\documentclass[preprint2]{aastex}
\shorttitle{Spectral variability of ULXs}
\shortauthors{Mizuno et al.}

\begin{document}

\title{Spectral variability of Ultra Luminous Compact X-ray Sources \\
in Nearby Spiral Galaxies}

%\author{T. Mizuno\altaffilmark{1}}
\author{T. Mizuno}
\affil{Department of Physics, Hiroshima University,
1--3--1 Kagamiyama, Higasho-Hiroshima, Hiroshima, Japan 739-8526}
\email{mizuno@hirax6.hepl.hiroshima-u.ac.jp}

\author{A. Kubota and K. Makishima}
\affil{Department of Physics, University of Tokyo,
7--3--1 Hongo, Bunkyo-ku, Tokyo, Japan 113-0033}

%\altaffiltext{1}{present address:???}

\begin{abstract}
Using the X-ray data taken with {\it ASCA}, a detailed analysis was made
of intensity and spectral variations of three ultra-luminous
extra-galactic compact X-ray sources (ULXs);
IC~342 source~1, M81~X-6, and NGC~1313 source~B,
all exhibiting X-ray luminosity in the range
$2 \times 10^{39}$--$1.5 \times 10^{40}$~erg~s$^{-1}$.
As already reported, IC~342 source~1 showed short-term 
X-ray intensity variability by a factor of 2.0 on a typical time scale
of 10~ks.
M81~X-6 varied by a factor of 1.6 across seven observations
spanning 3 years, while NGC~1313 source~B varied by a factor
of 2.5 between two observations conducted in 1993 July
and 1995 November.
The $ASCA$ spectra of these sources, acquired on these occasions,
were all described successfully as optically-thick
emission from standard accretion disks around black holes.
This confirms previous $ASCA$ works which explained
ULXs as mass-accreting massive black-hole binaries.
In all three sources, the disk color temperature was uncomfortably high
at $T_{\rm in}=1.0$--2.0~keV,
and was found to vary in proportion to the square-root of the
source flux.
The apparent accretion-disk radius is hence inferred to change
as inversely proportional to $T_{\rm in}$.
This suggests a significant effect of advection
in the accretion disk.
However, even taking this effect fully into account,
the too high values of $T_{\rm in}$ of ULXs cannot be
explained.
Further invoking the rapid black-hole rotation may give
a solution to this issue.
\end{abstract}


\section{Introduction}

One of the most puzzling aspects of X-ray emission
from normal galaxies is the presence
of exotic, powerful, apparently point-like
X-ray sources in arm regions of some spiral galaxies
%(e.g., Fabbiano 1989).
\citep{Fabbiano1989}.
We call them ``ultra-luminous compact X-ray sources (ULXs)'',
because their luminosities,
typically 10$^{39\mbox{--}40}~{\rm erg~s^{-1}}$,
are one to two orders of magnitude
higher than those of the most luminous Galactic X-ray binaries.
Since the Eddington limit for an object with mass $M$
is expressed as
\begin{equation}
L_{\rm E} =1.5 \times 10^{38}
\left(
\frac{M}{M_{\odot}}
\right)
~{\rm erg~s^{-1}}~~~,
\label{eqn:eddington}
\end{equation}
where $M_{\odot}$ denotes the solar mass,
ULXs were regarded as mass accreting black-hole binaries (BHBs)
involving massive
(50--100~$M_{\odot}$) stellar-mass black holes (BHs),
or neutron-star binaries (NSBs) with highly collimated X-ray emission.
However, both explanations remained speculative, 
because X-ray spectra of ULXs were poorly known.

The breakthrough has been brought by {\it ASCA}
%(Tanaka et al.\ 1994),
\citep{Tanaka1994},
thanks to its fine spectral capability
with a modest angular resolution.
Following previous works on individual ULXs
(e.g., IC~342 source~1; Okada et al.\ 1998),
%Makishima et al.\ (2000)
\citet{Makishima2000}
have performed an 
extensive spectral
investigation of ULXs,
and found that their spectra are commonly represented
by a so-called multi-color disk-blackbody
(MCD) model
that describes emission from an accretion disk
around a compact star
%(e.g., Mitsuda et al.\ 1984, Makishima et al.\ 1986),
\citep{Mitsuda1984,Makishima1986},
strongly reinforcing the BH interpretation.
The MCD model provides two basic quantities of the 
accretion disk;
the inner-most disk temperature, $T_{\rm in}$, and the
inner-most disk radius, $R_{\rm in}$.
Comparing these two quantities of ULXs with those
of Galactic/Magellanic BHBs,
%Makishima et al.\ (2000)
\citet{Makishima2000}
pointed out that 
the obtained values of $T_{\rm in}$
are significantly higher
than those of ordinary BHBs,
whereas the values of 
$R_{\rm in}$ are not so much different.
Since $R_{\rm in}$ might 
correspond to the last stable orbit around a
central BH,
%(e.g., Ebisawa 1991, Ebisawa et al.\ 1993),
%\citep{Ebisawa1991, Ebisawa1993},
this fact implies that the mass of BHs in ULXs
are relatively low ($\sim$10--20~$M_{\odot}$),
being inconsistent with their high luminosities.
To cope with this issue, 
they suggested that
the central BH in a ULX is rapidly rotating
(i.e., a Kerr BH),
hence
the disk can get closer to the 
BH and the observed value of $T_{\rm in}$
can become higher.
%This idea was originally proposed by Zhang et al.\ (1997)
This idea was originally proposed by \citet{Zhang1997}
to explain
the similar high values of $T_{\rm in}$ found from the two
Galactic jet sources, GRO~J1655-40 and GRS~1915+105.

Although our understanding of ULXs have thus made a great progress,
previous works
have been dealing only with the
time-averaged spectra, 
and left their spectral variability
unexamined.
Such a study would bring us key information on the source nature.
For example, $Ginga$ observations of variable Galactic/Magellanic BHBs
%(Ebisawa 1991, Ebisawa et al.\ 1993)
\citep{Ebisawa1991, Ebisawa1993}
revealed that their source luminosities vary as
$\propto T_{\rm in}^{4}$, or $R_{\rm in}$ remains constant;
this fact implies
that an optically-thick, geometrically-thin
standard accretion disk
%(Shakura \& Sunyaev\ 1973)
\citep{S-S1973}
is actually realized around a Schwarzschild BH,
where the inner disk radius is defined by the
last stable circular orbit.
We also point out that in the ``Kerr BHBs'' scenario of ULXs,
only a decrease of the inner disk radius has been considered.
Needless to say, it is also necessary to examine how the
relativistic effects (i.e., gravitational redshift,
Doppler shifts, and the light bending)
affect the observed spectra.

Accordingly, we performed (re-)analysis of the 
spectra of three
luminous ULXs; IC~342 source~1, M81 X-6, and NGC~1313 source~B.
The first has the largest flux among the observed ULXs next to
M33 X-8, and showed strong short-term variability 
%(e.g.\ Okada et al.\ 1998).
\citep{Okada1998}.
The other two have relatively high fluxes and have been observed 
several times with {\it ASCA},
and are hence suitable for the study on the spectral variability.
We describe the {\it ASCA} observation
of these sources in the next section,
and examine how their spectra varied in \S~3.
We discuss their source natures in \S~4.

\section{Targets and Observations}

IC~342 is a nearby starburst Scd galaxy located
close to the Galactic plane
($b \sim 10^{\circ}$).
%After Tully\ (1988), we assume the distance as 3.9~Mpc,
After \citet{Tully1988}, we assume the distance as 3.9~Mpc,
although large optical extinction caused by the low Galactic
latitude makes the distance uncertainty relatively large.
This galaxy has been observed by
{\it ASCA} twice;
in 1993 September and 2000 Feburuary.
%As already described in Okada et al.\ (1998),
As already described in \citet{Okada1998},
source~1 in this galaxy exhibited a significant variability
during the 1993 observation (see also Figure \ref{fig:ic342 lc}).
Accordingly,
we investigate the change of its spectral property
in detail in \S~3.1.
The spectral difference between the 1993 and 2000
observations is described in a separate paper
\citep{Kubota2001}.
%(Kubota et al.\ 2000).

M81 (NGC~3031) has a Cepheid-based accurate
distance of 3.6~Mpc from the 
%Hubble Space Telecsope  observations (Freedman et al.\ 1994).
Hubble Space Telecsope  observations \citep{Freedman1994}.
%Fabbiano (1988)
\citet{Fabbiano1988}
observed this galaxy with $Einstein$ and 
detected several sources, including the brightest one, called X-5
%(see also Ishisaki et al.\ 1996 and Iyomoto 1999)
\citep{Ishisaki1996, Iyomoto1999}
at the nucleus,
and the most luminous off-nucleus source called X-6
which we study in the present paper.
Due to the explosion of SN~1993J, $ASCA$ has frequently observed
the M81 region
%(Kohmura et al.\ 1994; Kohmura 1994; Uno 1997).
\citep{KohmuraPHD1994, Kohmura1994, Uno1997}.
Among these observations, we utilize the data obtained after 1994,
in which SN~1993J had faded away significantly and its contamination
to the X-6 spectrum is sufficiently low.
Thus we analyzed data from seven observations in total, as shown in
Table~\ref{tab:sample}.
Although the first two datasets were already investigated by
%Makishima et al.\ (2000),
\citet{Makishima2000},
we here deal with them as well as
other five,
in order to study the long-term variability.
The same seven datasets were also 
employed by
%Uno (1997),
\citet{Uno1997},
in his study of SN~1993J.

NGC~1313 is a nearby face-on, late-type Sb galaxy at a
distance of 4.5~Mpc
%(Vaucouleurs 1963).
\citep{Vaucouleurs1963}.
X-ray observations using $Einstein$~$Observatory$, $ROSAT$, and $ASCA$
%(Fabbiano \& Trinchieri\ 1987; Colbert et al.\ 1995;
%Miller et al.\ 1998; Petre et al.\ 1994)
\citep{Fabbiano1987, Colbert1995, Miller1998, Petre1994}
showed that its X-ray emission is dominated by
three extremely luminous point-like sources of
$L_{\rm X}$ $\sim$ 10$^{39}$~erg~s$^{-1}$ each.
One of them is an X-ray luminous supernova, namely SN~1978K,
and the other two are ULXs called source~A and source~B
after
%Petre et al. (1994).
\citet{Petre1994}.
The former is located close ($\sim$45$''$) to the galaxy nucleus,
whereas the latter is seen at the south-end of the host galaxy.
This galaxy has been observed by {\it ASCA} twice,
in 1993 July and in 1995 November.
%Makishima et al.\ (2000)
\citet{Makishima2000}
studied source~B,
and showed that the spectrum of this source 
on either occasion
can be expressed by the MCD model with different
values of $T_{\rm in}$.
We here investigate this ULX
for both observations in \S~3.3,
concentrating on the spectral variability.
As for source~A, the two observations will be compared in another paper 
(T. Mizuno et al.\ in preparation).

%X-ray images of all our three sample galaxies are shown in
%Figure~\ref{fig:image}, superposed on the optical ones.

The observational log of our target sources
is summarized in Table~\ref{tab:sample}.
For M81~X-6, we did not use the GIS data (see \S~3.2).
The GIS data for NGC~1313 in the second observation was acquired
in a bit assignment (PH--X--Y--RT--SP--Time) of 8--8--8--0--0--7;
this is an exceptional mode which sacrifices rise time information
to improve the time resolution (in search for a pulser in SN~1978K),
so that we did not apply the off-line rise-time cut
screening on the data.
For the other GIS data and the SIS data,
we applied a standard data screening and tabulated the net exposure
in Table~\ref{tab:sample}.

\placetable{tab:sample}

\section{Data Analysis}

\subsection{IC~342 source~1}

The time-averaged spectra of IC~342 source~1
were already analyzed by
%Okada et al.\ (1998)
\citet{Okada1998}
and
%Makishima et al.\ (2000).
\citet{Makishima2000}.
They accumulated the on-source spectra from a circular region
of 3$'$ radius
on the source center, subtracted the background
spectra extracted from blank-sky observations, and fitted the obtained
spectra with single component models.
According to their results,
the power-law model is completely unacceptable,
the thermal bremsstrahlung (TBS) model
provides much better value of $\chi^{2}$ but
is also rejected at 99\% confidence level,
whereas an MCD one gives an acceptable fit.

\begin{figure}
\plottwo{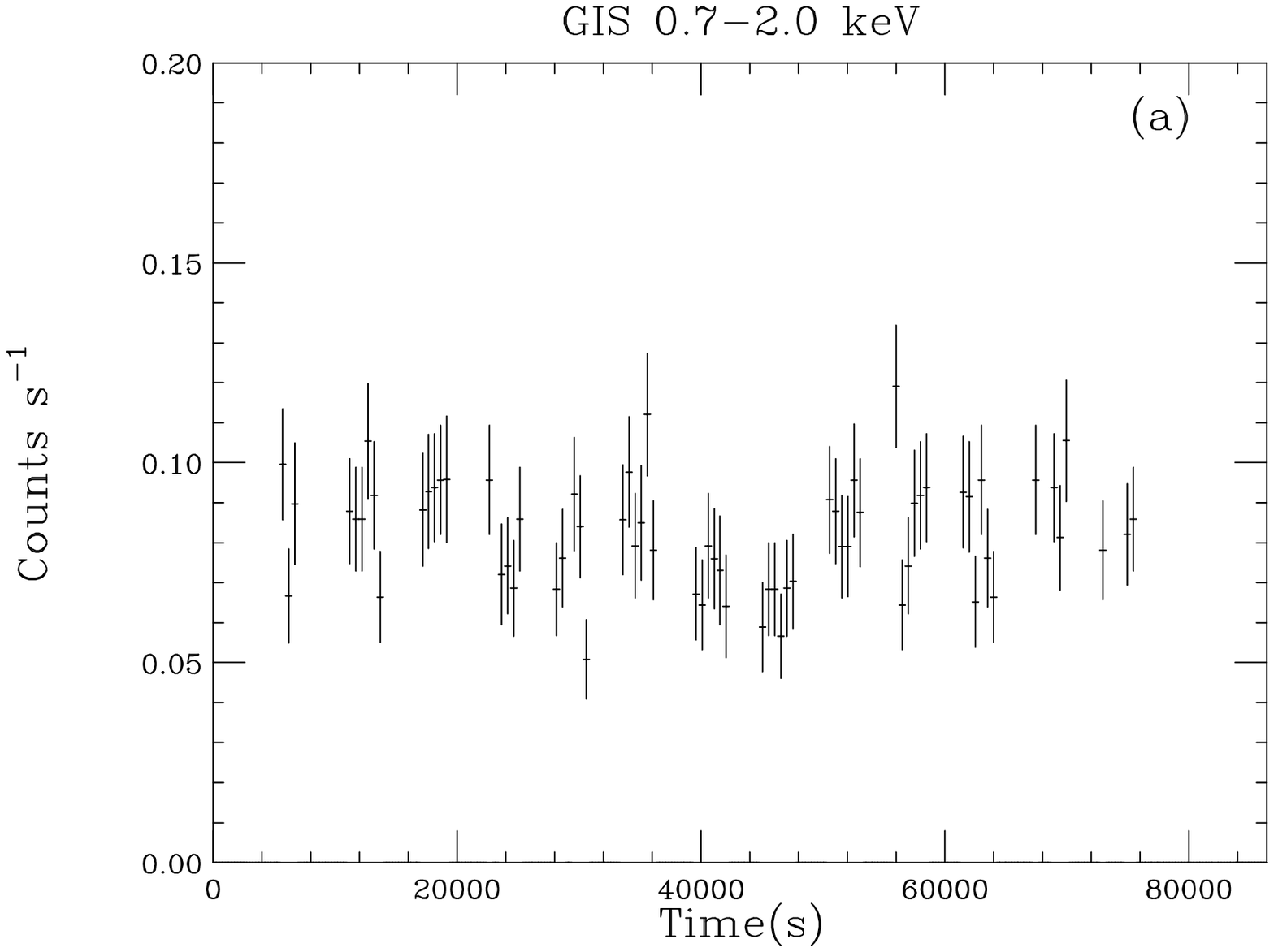}{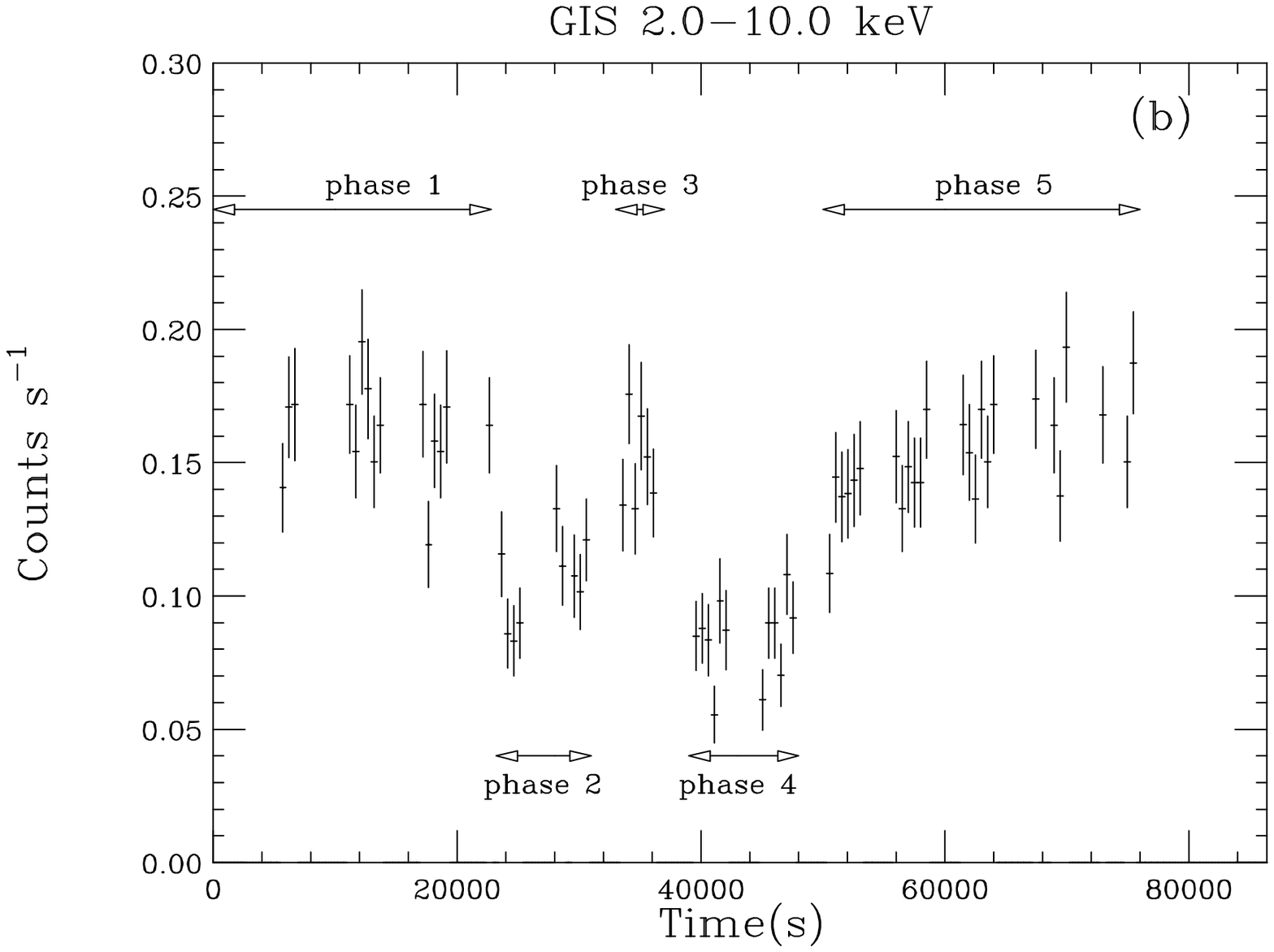}
%\plottwo{low.ps}{high.ps}
%\plotone{low.ps}
%\plotone{high.ps}
\caption{The {\it ASCA} GIS2 + GIS3 light curve of IC~342 source~1
including background.
Panel (a) represents the light curve in the energy range
of 0.7--2.0~keV,
whereas panel (b) in 2.0--10~keV.
The background count rate for the low and high energy range
is about 0.004 c~s$^{-1}$ and 0.006 c~s$^{-1}$, respectively.
\label{fig:ic342 lc}
}
\end{figure}

\begin{figure}
%\plotone{ic342_spec.eps}
\plotone{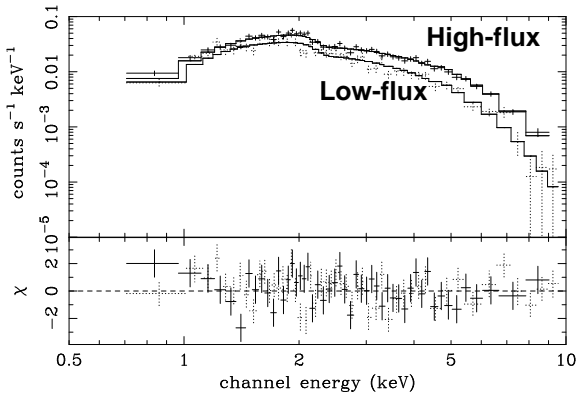}
\caption{The high and the low flux phase spectra of IC~342 source~1,
fitted with the MCD model.
The histograms show the best fit model and the crosses represent
the observed spectra.
The lower panel indicates the fit residuals.
Although the fit is simultaneous to the SIS and the GIS spectra,
here only the GIS spectra are shown for clarity.
\label{fig:ic342s1_time}
}
\end{figure}

Although the short-term variability of source~1 was already sketched
by 
%Makishima (1994)
\citet{Makishima1994}
and
%Okada et al.\ (1998),
\citet{Okada1998},
we reproduced
its GIS light curves in Figure~\ref{fig:ic342 lc}.
As already reported by
%Makishima (1994)
\citet{Makishima1994}
and
%Okada et al.\ (1998),
\citet{Okada1998},
the source exhibits clear time variability by a factor of 2
in a few hours,
especially in the hard energy band.
In order to study this behavior in more detail,
we divided both the SIS and the GIS data
into five time regions as defined in Figure~\ref{fig:ic342 lc}b,
and analyzed the corresponding time-sorted spectra.
To grasp the rough information of the spectral variability,
we first summed the phase 2 and 4 spectra into
a ``low-flux phase'' spectra,
and summed the phase 1, 3, and 5 spectra into 
``high-flux phase'' spectra, as shown in Figure~\ref{fig:ic342s1_time}.
In agreement with the light curves,
the difference between the two spectra is more significant at 
higher energies, indicating that the spectrum hardens
as the source flux increases.
We fitted the high-flux phase SIS/GIS spectra jointly
with the MCD model, and obtained the result
as given in Table~\ref{tab:fit_ulxs};
the fit turned out to be acceptable ($\chi/\nu = 129.0/106$).
We also fitted the low-flux phase spectra;
the fit is again acceptable ($\chi/\nu = 115.6/107$).
The best fit values of the line-of-sight absorption 
are $N_{\rm H} = 4.7 \times 10^{21}~{\rm cm}^{-2}$
for both spectrua,
and are consistent with the time-averaged
one obtained by
%Makishima et al (2000).
\citet{Makishima2000}.
We also found that the disk temperature differs by a factor of 1.3
between two phases,
whereas the source flux by a factor of 1.7

In order to examine changes in the physical
condition of the accretion disk,
we calculated the values of $R_{\rm in}$
for the spectra of high-flux and low-flux phases.
Assuming a face-on geometry,
they became $112 \pm 8$~km and $149 \pm 19$~km, respectively,
where we adopted
the ratio of a color temperature
to an effective temperature of 1.7
after
%Shimura \& Takahara (1995),
\citet{S-T1995},
and applied a correction for
the inner-disk boundary condition after
%Kubota et al (1998).
\citet{Kubota1998}.
To our surprise,
the radius $R_{\rm in}$ does not appear to be constant
but increases as the flux decreases;
this behavior contradicts that of ordinary BHBs
%(e.g., Ebisawa\ 1991, Ebisawa et al.\ 1993).
\citep{Ebisawa1991, Ebisawa1993}.
To investigate this inference more quantitatively,
we fitted the two sets of spectra simultaneously
by constraining $N_{\rm H}$ and $R_{\rm in}$ to take the same values
between them (while allowing $T_{\rm in}$ to vary independently),
and then letting $R_{\rm in}$ to change separately.
Then, the fit improved from $\chi^{2}/\nu = 266.4/216$ to
$\chi^{2}/\nu = 245.0/215$, indicating that the change of $R_{\rm in}$
is statistically
significant (at 99\% confidence level by an $F$-test).

Although $R_{\rm in}$ thus changes 
significantly,
it might be an artifact due to wrong modeling
of the spectra. For example, 
the MCD emission from NSBs is
usually accompanied by a black-body hard component,
and BHBs in the soft state often exhibit a spectral
hard-tail in addition to the MCD component.
Fluctuation of such a hard component,
if any, might apparently cause the correlated changes of
$T_{\rm in}$ and $R_{\rm in}$ seen from IC~342 source~1.
We therefore tried to fit the high/low-flux phase spectra simultaneously
by a common MCD model with the same $R_{\rm in}$ and $T_{\rm in}$,
adding a hard-component model to the high flux phase spectra only.
The hard component was modeled by a 
black-body of temperature fixed at 2.0~keV,
typical for NSBs
%(e.g., Mitsuda et al.\ 1984),
\citep{Mitsuda1984},
or a power-law of photon index
fixed at $\Gamma = 2.5$,
nominal value for BHBs in the soft state
%(Tanaka \& Lewin\ 1995).
\citep{Tanaka1995}.
Then, the MCD plus power-law model could not explain the observed
spectra ($\chi^{2}/\nu = 492.2/216$).
The MCD plus black-body model was
also statistically unacceptable
($\chi^{2}/\nu = 312.1/216$).
Thus, the observed spectral variability cannot be due to fluctuations
of the hard component, but should be attributed
to the change of the MCD component itself.

\begin{figure}
%\plotone{cont.eps}
\plotone{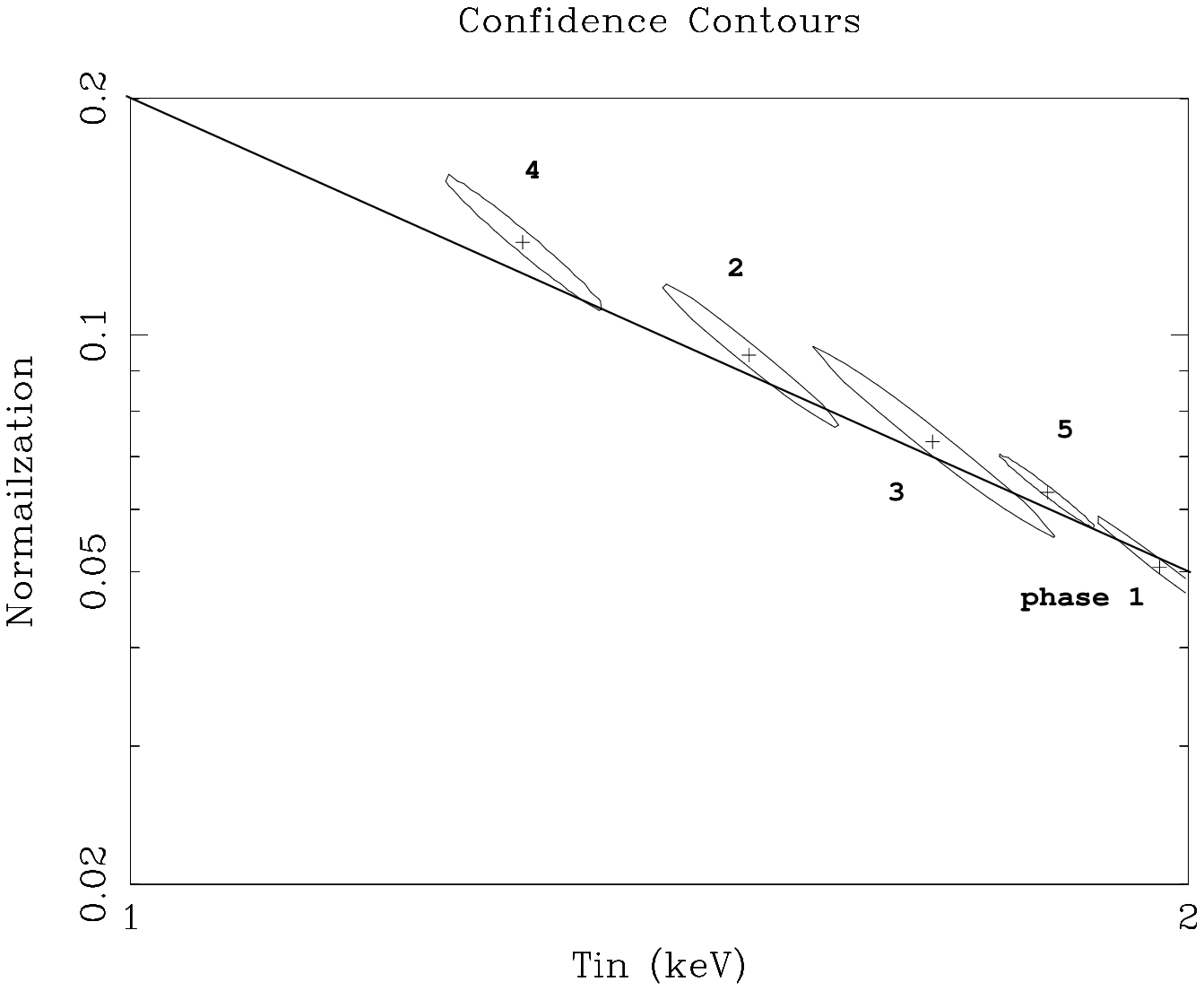}
\caption{The 68\% confidence contours of the short-term variations
of IC342 source~1, in terms of
$T_{\rm in}$ and the MCD normalization
(proportional to $R_{\rm in}^{2}$).
The solid line represents the relation
of $T_{\rm in} \propto R_{\rm in}^{-1}$.
The horizontal axis is logalithmic.
\label{fig:ic342s1_5_cont.eps}
}
\end{figure}

We finally fitted five time-sorted spectra 
for the five time intervals individually by the
MCD model, with the absorption fixed at
the best-fit value of high/low-flux phase spectra
($N_{\rm H} =4.7 \times 10^{21}$~cm$^{-2}$).
We plotted the confidence contours of the five spectra
on the $T_{\rm in}$--$R_{\rm in}^{2}$ plane
in Figure~\ref{fig:ic342s1_5_cont.eps},
where $R_{\rm in}^{2}$ is proportional to the normalization
of the MCD model.
Then, like we have already found in the two phase spectral fitting,
$R_{\rm in}$ gradually decreases as $T_{\rm in}$ increases,
and the relation is approximately expressed as
$R_{\rm in} \propto T_{\rm in}^{-1}$.
Further examination of this result will be discussed
in \S~4.

\placefigure{fig:ic342 lc}

\placefigure{fig:ic342s1_time}

\placefigure{fig:ic342s1_5_cont.eps}

\subsection{M81~X-6}

\subsubsection{Accumulation of the source and the background spectra}

When studying the spectrum of X-6, we should eliminate the contamination
from the two nearby sources,
SN~1993J and M81~X-5, the latter being the low-luminosity
active-galactic nucleus of M81
%(Ishisaki et al.\ 1996, Iyomoto 1999).
\citep{Ishisaki1996, Iyomoto1999}.
This is because 
X-6 is separated only $\sim$1$'$ from SN~1993J, and
$\sim$3$'$ from X-5,
as previously reported by
%Kohmura et al.\ (1994) and Ishisaki et al.\ (1996).
\citet{Kohmura1994} and \citet{Ishisaki1996}.

We accumulated the source spectrum of X-6 from a circular region
of 1.\hspace{-2pt}$'$5 radius,
to make the contamination from X-5 as low as possible.
For the same reason,
we only used the SIS data;
the poorer spatial resolution of the GIS
would increase the X-5 contamination.
In the obtained spectrum, nevertheless,
typically $\sim$50\% photons still originate from X-5
due to its brightness.
In order to remove this residual contamination, we accumulated a
background spectrum over another region
having the same size as was used for the on-source spectrum.
This background region is located opposite to X-6 with respect to X-5,
where we expect a similar amount of contamination from X-5.
We multiplied a constant factor to thus obtained background spectrum,
and then subtracted it from the on-source spectrum.
This ``scaling factor'' was introduced to take into account the
$ASCA$ XRT's asymmetric point spread function, and
was determined based on the ray-tracing (Monte-Carlo)
simulation developed by the XRT team.
By simulating the X-5 event for each observation
(each position of X-5 on the focal plane),
we can estimate the ratio of photons in the source region
to the background region.
The ratio, or the scaling factor, is typically 1.5, ranging over
1.2--1.8.
This procedure also subtracts the non X-ray background (NXB) and
the Cosmic X-ray background (CXB) after multiplying the same
scaling factor, which is not precise for the NXB and CXB.
However, we can neglect this effect, since the
ratio of the NXB + CXB count rate to that of 
the contaminated photons from M81~X-5 is estimated to be
$\le 10\%$.

\begin{figure}
%\plotone{../spec/m81x6_9404_diskbb_add_fit_rot.eps}
%\plotone{m81x6_9404_diskbb_add_fit_rot.eps}
\plotone{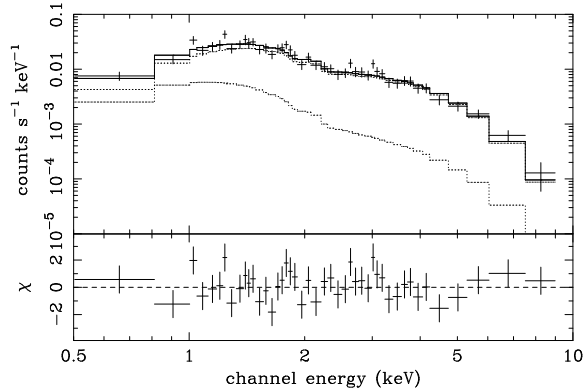}
\caption{
The SIS spectra of M81~X-6 obtained in 1994 April,
fitted with an MCD model. 
Dotted lines indicate individual contributions
of SN~1993J (lower one) and X-6 (higher one),
with the former represented by a fixed power-law component (see text).
\label{fig:m81x6}
}
\end{figure}

Thus, we have obtained the spectrum for X-6 plus SN~1993J.
The contamination from SN~1993J is difficult to eliminate
because of their short separation.
Instead, we take it into account in a different way.
The SN~1993J spectrum was separately estimated by
%Uno (1997)
\citet{Uno1997}
through one-dimensional SIS image analysis, which was
originally developed by
%Kohmura (1994).
\citet{KohmuraPHD1994}.
%According to Uno (1997),
According to \citet{Uno1997},
SN~1993J had faded away
significantly after one year from the explosion,
and the spectrum is expressed by a power-law of
$\Gamma=2.5^{+1.4}_{-0.8}$ and the X-ray flux
$f_{\rm X}$~=~0.5~$\pm$~0.2~$\times$~10~$^{-12}$~erg~s$^{-1}$~cm$^{-2}$
in 1994 April,
and of $\Gamma=3.0^{+2.0}_{-1.0}$ and
$f_{\rm X}$~=~0.2~$\pm$~0.1~$\times$~10~$^{-12}$~erg~s$^{-1}$~cm$^{-2}$
in 1994 October, both in the 0.5--8 keV band.
We calculated their contamination to the X-6 spectrum
based on the ray-tracing simulation, and took the results into account
as a fixed power-law component when fitting the X-6 spectrum,
as indicated in Figure~\ref{fig:m81x6}.
The contribution of SN~1993J to the 0.5--10 keV flux turned out to be
small; only $\sim$ 10\% and $\sim$ 3\% in 1994 April and October,
respectively.
For the other data which were obtained after 1995 April,
we neglected the contribution from SN~1993J, since the supernova
further faded away.

\subsubsection{Spectra of individual observations}

We fitted the obtained seven spectra separately
with the typical single component models;
the power-law, the TBS, and the MCD model.
Then, the MCD model turned out to be always acceptable at
90\% confidence level, while the TBS model provided a worse fit
except for the 1995 October and 1996 October spectra
(for the worst case, $\chi^{2}/\nu = 69.8/54$ in 1996 April).
The power-law
fit gave the largest values of $\chi^{2}$
(1996 April data gave the worst fit of $\chi^{2}/\nu = 80.3/54$).
Therefore we conclude that the spectra of M81 X-6,
like those of IC~342 source~1,
can be represented well by the single MCD model,
and tabulate the MCD-fit results in Table~\ref{tab:fit_ulxs}.
Our results for the 1994 observation are consistent
with those previously reported by
%Makishima et al.\ (2000).
\citet{Makishima2000}.

\begin{figure}
%\plotone{m81fxTin.ps}
\plotone{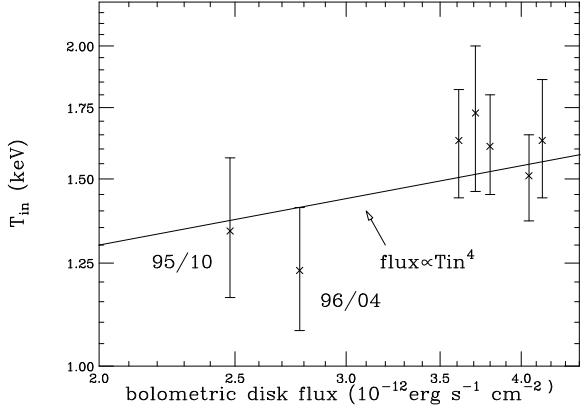}
\caption{Relation between the disk bolometric flux
and $T_{\rm in}$ for M81~X-6.
\label{fig:m81fxTin}
}
\end{figure}

We plotted the obtained values of the bolometric
flux of the accretion disk
$f_{\rm bol}^{\rm disk}$ and $T_{\rm in}$
in Figure~\ref{fig:m81fxTin}.
It indicates that the seven datasets can be grossly grouped into
two representative states;
a low-temperature state observed in 1995 October and 1996 April,
and a high-temperature state observed on the other occasions.
The low/high-temperature states correspond to the low/high-flux states,
so that M81~X-6 shows similar spectral variability to
IC~342 source~1.
We also present the line of
$f_{\rm bol}^{\rm disk} \propto T_{\rm in}^{4}$,
i.e., the locus of $R_{\rm in}$ remaining constant.
Although the statistical errors are somewhat large, a slight
deviation from this line can be inferred.

\begin{figure}
%\plottwo{m81x6_cont_edit.eps}{m81x6_hl_cont_edit.eps}
\plottwo{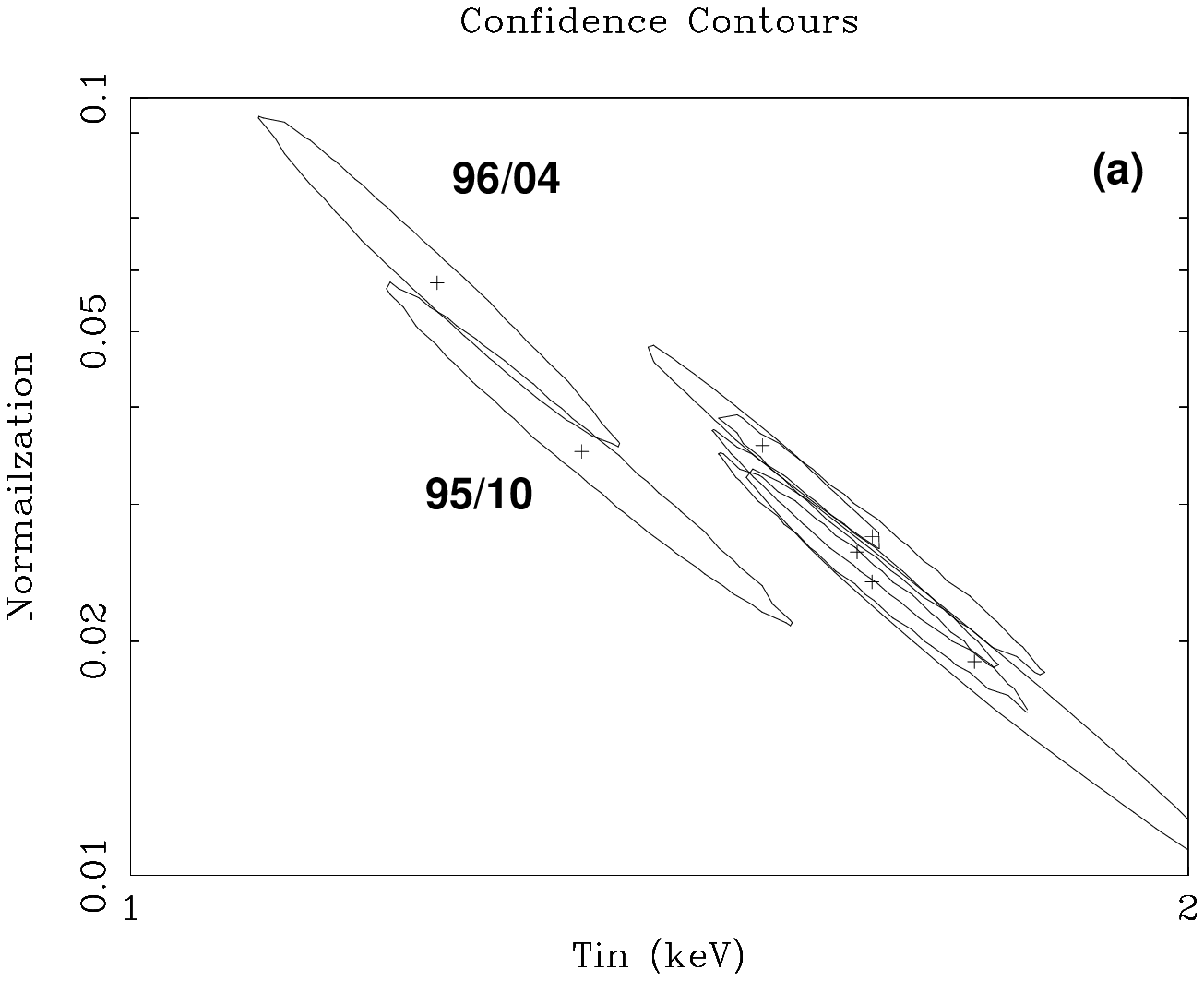}{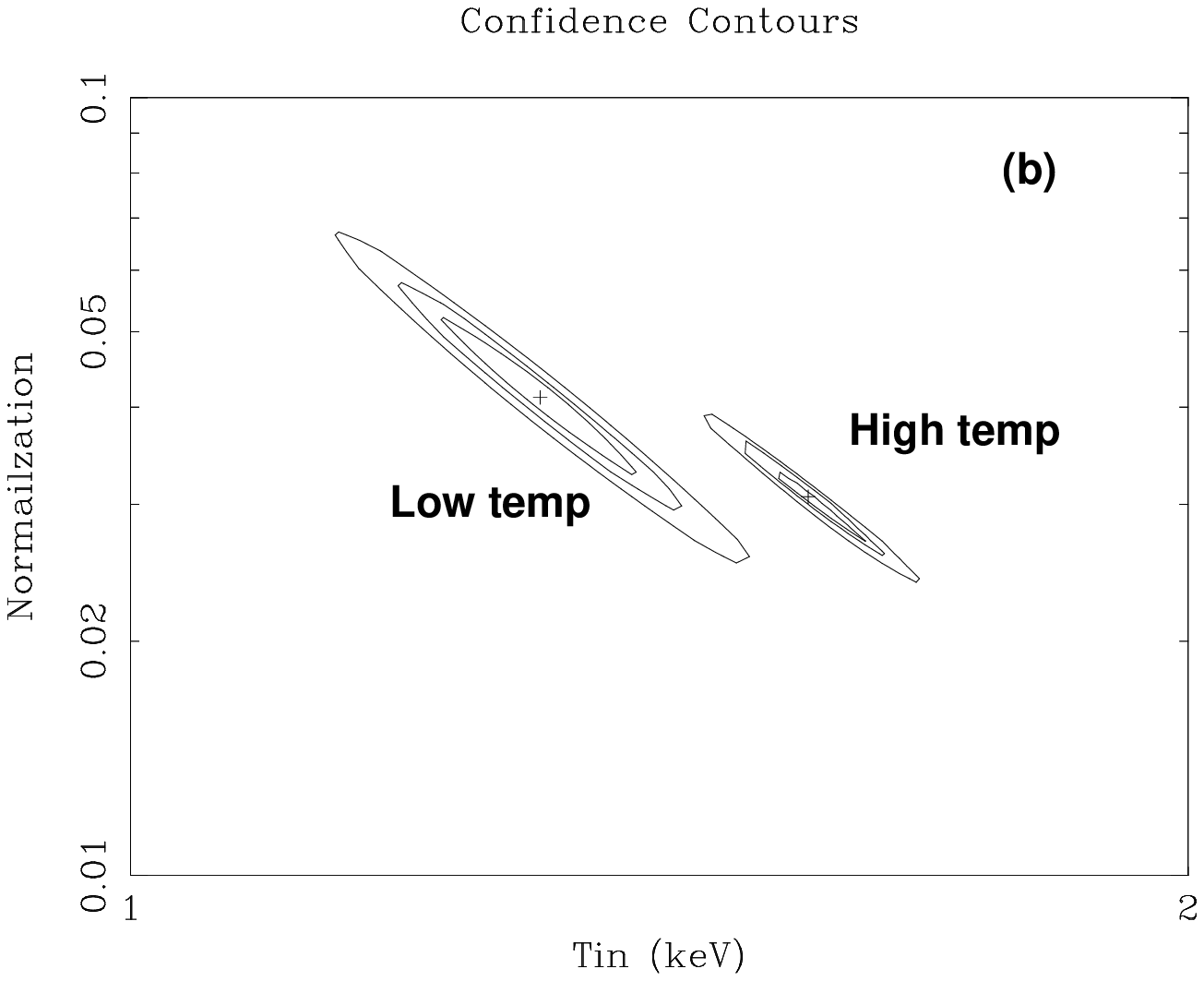}
%\plotone{m81x6_cont_edit.eps}
%\plotone{m81x6_hl_cont_edit.eps}
\caption{The same as
Figure \protect\ref{fig:ic342s1_5_cont.eps},
but for M81~X-6 instead of IC~342 source~1.
(a) The contours of individual observations,
where only the 68\% confidence levels are shown.
(b) The contours of the high/low-temperature state spectra
obtained by grouping the individual datasets.
The 68\%, 90\%, and 99\% confidence levels are presented.
\label{fig:m81x6 cont}
}
\end{figure}

We hence made 
the $T_{\rm in}$--$R_{\rm in}^{2}$ diagram
in Figure \ref{fig:m81x6 cont}a,
as was done on IC~342 source~1.
%However we do not necessarily mean that the variation
%of X-6 is bimodal.
Thus, we again, find a hint of anti-correlation between
$T_{\rm in}$ and $R_{\rm in}^{2}$, like in the case of
IC~342 source~1,
but relatively large errors hampers definite statement.
We therefore grouped the seven datasets into two spectra, 
i.e., the high-temperature state spectrum
and the low-temperature state one.
We fitted these two spectra by the same single component models
as were used for each observation,
and, again, the MCD model turned out to be the best
representation of the data.
We summarized the 
best fit parameters of the MCD model
in Table~\ref{tab:fit_ulxs},
and plotted the $T_{\rm in}$--$R_{\rm in}^{2}$ diagram in
Figure~\ref{fig:m81x6 cont}b,
by fixing the absorption at the average value of the two states
($N_{\rm H} = 1.8 \times 10^{21}$~cm$^{-2}$).
Thus,
$R_{\rm in}$ increases marginally as the flux decreases,
although not so noticeable as in the case of IC~342 source~1.
In fact, when we fitted the two subgrouped spectra simultaneously
by constraining $N_{\rm H}$ and $R_{\rm in}$ to be common,
we obtained an acceptable fit with $\chi^{2}/\nu$=140.4/126,
while letting $R_{\rm in}$ to be free
does not improve the fit significantly ($\chi^{2}/\nu$=138.0/125).
Therefore $R_{\rm in}$ is consistent with being the same
between the two spectra, although a weak anti-correlation
between $T_{\rm in}$ and $R_{\rm in}$ may be
present.

\placefigure{fig:m81x6}

\placefigure{fig:m81fxTin}

\placefigure{fig:m81x6 cont}

\subsection{NGC~1313 source~B}

As previously reported by
%Makishima et al.\ (2000),
\citet{Makishima2000},
this source showed time variability by a factor of $\sim 2.5$
between the 1993 and the 1995 observations.
They analyzed both spectra with single component models
(the power-law, the TBS, and the MCD model),
and found that the MCD one gives the best description of the data.
Moreover, they pointed out that the disk temperature is positively
correlated with the source flux, and the value of $R_{\rm in}$ seemed
to increase for two years
whereas the source flux decreased.
This behavior is therefore similar to that found
with the two sources
we have described so far,
and we investigate the variability of this ULX in further detail.

\begin{figure}
%\plotone{../fig_ulxs2/n1313sb_cont_edit.eps}
%\plotone{n1313sb_cont_edit.eps}
\plotone{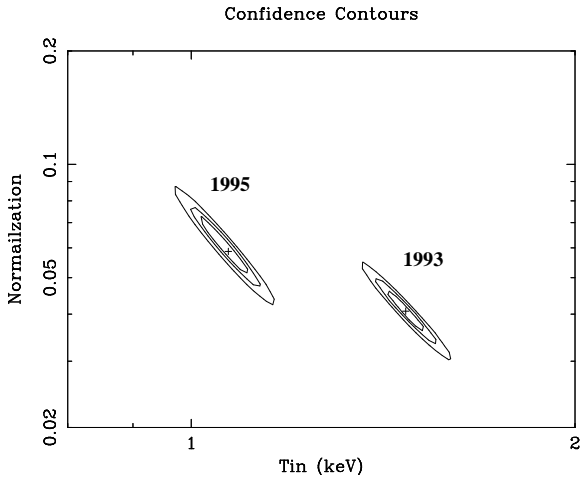}
\caption{The same as
Figure \protect\ref{fig:ic342s1_5_cont.eps} and
\protect\ref{fig:m81x6 cont},
but for the NGC~1313 source~B, where 68\%, 90\%, and 99\%
confidence levels are shown.
\label{fig:n1313sb_cont_edit.eps}
}
\end{figure}

We here utilized the same data as used by
%Makishima et al.\ (2000),
\citet{Makishima2000},
and fitted them with the MCD model.
We plot the confidence contours on the $T_{\rm in}$-$R_{\rm in}^{2}$
plane in Figure~\ref{fig:n1313sb_cont_edit.eps},
where we fixed the absorption at $N_{\rm H} = 7 \times 10^{20}$~cm$^{-2}$
(the average value between the two observations).
Like the two variable sources so far studied,
a slight increase in $R_{\rm in}$ can be seen as the flux decreases.
The change of $R_{\rm in}$ is statistically significant 
(at 90\% confidence);
we obtained an improved fit
($\chi^{2}/\nu$=214.9/207)
when fitting the two observations simultaneously allowing
$R_{\rm in}$ to take separate values,
compared with that ($\chi^{2}/\nu$=222.9/208)
obtained when constraining $R_{\rm in}$ to be common.

Since the source positions on the focal plane
are largely separated ($\sim 6^{'}$) between the two observations,
the obtained value of $R_{\rm in}$ might be affected
by the response uncertainty.
According to an extensive investigation utilizing the Crab Nebula
by
%Fukazawa et al.\ (1997),
\citet{Fukazawa1997},
the calculated nominal source flux may increase artificially
as the off-axis angle increases.
As for our two observations, source~B was observed at an 
off-axis angle of $\sim 10^{'}$ and $\sim 5^{'}$
in 1993 and 1995, respectively.
Therefore, if this residual response uncertainty is present, the true
1993 flux should be lower than the estimated value.
This would further enhance the increase in $R_{\rm in}$ for two years.
We therefore conclude that the change of $R_{\rm in}$
seen for NGC~1313 source~B
is a real effect rather than an artifact.

\placefigure{fig:n1313sb_cont_edit.eps}

\section{Discussion}

\subsection{Two distinctive properties of ULXs}

\begin{figure}
%\plotone{../summary/rin.ps}
%\plotone{rin.ps}
\plotone{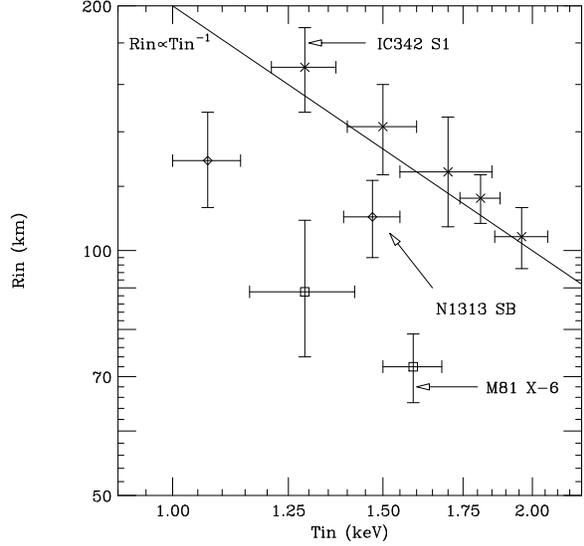}
\caption{Relation between $R_{\rm in}$ and $T_{\rm in}$ for 
the three ULXs.
The error bars represent the 90\% confidence errors 
of the spectral fit, and the solid line
indicates the relation of $R_{\rm in} \propto T_{\rm in}^{-1}$.
\label{fig:rin.ps}
}
\end{figure}

So far, we have been investigating the time variability of ULXs spectra,
and found that they can always be represented by the MCD model,
event when the source flux varies significantly.
We also found that
the observed flux is positively correlated
with $T_{\rm in}$.
If $R_{\rm in}$ remains constant as the source flux varies
like ordinary Galactic/Magellanic BHBs
%(e.g., Ebisawa\ 1991, Ebisawa et al.\ 1993, Kubota\ 2001),
\citep{Ebisawa1991, Ebisawa1993, KubotaPHD2001},
we will have the relation of $f_{\rm bol}^{\rm disk} \propto T_{\rm in}^{4}$,
whereas our ULXs show somewhat different relation
(e.g., Figure~\ref{fig:m81fxTin}).
As summarized in Figure~\ref{fig:rin.ps},
they seem to obey
a single common scaling of $R_{\rm in} \propto T_{\rm in}^{-1}$.

The observed values of $T_{\rm in}$ of our ULXs,
ranging 1.0--2.0 keV, are also different from those of
normal BHBs (typically 0.5--1.2 keV; e.g. Tanaka and Lewin 1995).
Such a contradiction has been known to be a common feature of ULXs
%(e.g., Okada et al.\ 1998, Mizuno et al.\ 1999, Makishima et al.\ 2000).
\citep{Okada1998, Mizuno1999, Makishima2000}.
As previously described by
%Makishima et al. (2000),
\citet{Makishima2000},
in the MCD approximation for the standard accretion disk
around a Schwartzshild BH of mass $M$,
$T_{\rm in}$ is
expressed as
\begin{equation}
T_{\rm in} = 1.2 \left( \frac{\xi}{0.412} \right)^{1/2}
\left( \frac{\kappa}{1.7} \right) \eta^{1/4}
\left( \frac{M}{10M_{\odot}} \right)^{-1/4} {\rm keV}~~~,
\label{eqn:T_Mass}
\end{equation}
where $\eta$ denotes the disk bolometric luminosity normalized
to the Eddington luminosity,
and we have been assuming $\xi=0.412$ and $\kappa=1.7$.
Thus, a heavier BH should show a lower value of
$T_{\rm in}$, as expressed by
$T_{\rm in} \propto M^{-1/4}$.
Nevertheless, 
the ULXs that have high luminosities and
hence high BH mass actually exhibit higher disk temperatures.
The upper-limit on $T_{\rm in}$
inferred from equation~\ref{eqn:T_Mass} is
0.68~keV for the phase 1 spectra of IC342 source~1,
1.0~keV for the high-temperature phase spectra of M81~X-6,
and 0.89~keV for the 1993 spectra of NGC~1313 source~B
even assuming a face-on geometry;
these predictions contradicts the observed values by a factor of
2.9, 1.6, and 1.7, respectively.
A higher disk inclination angle makes the situation even worse.
Even if we exclude IC~342 source~1, for which the distance
to the host galaxy is somewhat uncertain,
the contradiction by a factor of $\sim \sqrt{3}$ remains to be
solved.

Thus, our research with {\it ASCA} has provided ambivalent results
on the interpretation of ULXs in terms of optically-thick standard
accretion disks around massive stellar-mass BHs.
On one hand, we have confirmed that the MCD model remains
a good representation of the ULX spectra even when the source
flux varied considerably.
On the other hand, the innermost disk radius apparently
changes in an anti-correlation with the
disk temperature, which makes another discrepancy in addition to
the previously pointed-out problem of too-high values of
$T_{\rm in}$
%(e.g., Makishima et al.\ 2000).
\citep{Makishima2000}.
In order to solve these two problems, we may introduce
some modification to the standard disk picture.

\placetable{tab:fit_ulxs}

\placefigure{fig:rin.ps}

\subsection{Slim disk scenario}

Although the standard disk model successfully explains
the spectra from BHBs in the soft state,
a significant progress has been achieved on the theory of accretion
disks.
When accretion rate $\dot{M}$ is high and the
source luminosity approaches $L_{\rm E}$,
the standard disk is predicted to change into
a so-called optically-thick, advection-dominated accretion flow
(ADAF).
This ``optically-thick ADAF'' model
%(e.g., Abramowicz et al.\ 1988;
%Szuszkiewicz et al.\ 1996;
%Watarai et al.\ 2000)
\citep{Ab1988, Szu1996, Watarai2000}
is also named a ``slim accretion disk model'',
since it is moderately geometrically-thick.
In ULXs, because of their high luminosities,
the accretion flow configuration is expected to change from
the standard disk to the slim disk.

One of the most characteristic features of the slim disk,
revealed by 
%Watarai et al.\ (2000)
\citet{Watarai2000}
through their numerical calculations,
is that 
the X-rays are radiated
not only from the regions outside the last stable orbit,
but also from the regions inside it,
since an abrupt change of the radial infall velocity does
no longer occur.
They fitted the numerically calculated
slim disk spectrum by the MCD model
(with $R_{\rm in}$ and $T_{\rm in}$ being
the model parameters),
and found that
as $\dot{M}$ increases,
$R_{\rm in}$ decreases
as $R_{\rm in} \propto T_{\rm in}^{-1}$.
Thus, our finding of the change in $R_{\rm in}$
can be explained naturally
by presuming that the slim disk is realized in ULXs.
This suggests the presence of a slim disk in the three
ULXs studied here.

Because the slim disk has a smaller $R_{\rm in}$
and a higher $T_{\rm in}$
than a standard disk in the same condition,
it could also explain the ``too-high $T_{\rm in}$'' problem
described in \S~4.1.
For the present three ULXs,
the slim-disk scaling
(the bolometric luminosity
$L_{\rm bol} \propto R_{\rm in}^{2} T_{\rm in}^{4} \propto T_{\rm in}^{2} $)
holds over a typical range
from the observed highest luminosity $L_{\rm max}$
(which we tentatively identify with $L_{\rm E}$)
down to $\sim L_{\rm max}/2$.
While the source luminosity varies by a factor of 2,
we expect the disk
temperature to
change by a factor of $2^{1/4}$ for the standard disk,
or $2^{1/2}$ for the slim disk.
In this way, the slim disk scenario can relax the
too-high $T_{\rm in}$ problem at least by a factor of 
$\frac{2^{1/2}}{2^{1/4}} \sim 1.2$.
However, in order to fully resolve the factor $\sqrt{3}$ discrepancy
seen for our three ULXs,
the source must make a transition from the standard-disk
to slim-disk regimes at still lower luminosities, e.g.,
$L_{\rm bol} = \frac{1}{\sqrt{3}^4}L_{\rm E} \sim 0.1~L_{\rm E}$.
This contradicts the observed results
of Galactic/Magellanic BHBs,
where the standard accretion disk picture has been confirmed to be valid
at least up to $\sim \frac{2}{3}~L_{\rm E}$ in case of GS~2000+25 and
LMC~X-3
%(e.g., Ebisawa\ 1991, Ebisawa et al.\ 1993,
%Mineshige et al.\ 1994, Makishima et al.\ 2000). 
\citep{Ebisawa1991, Ebisawa1993,
Mineshige1994, Makishima2000}.

Through the present study, we suggest that the ULXs are
in the slim-disk condition because of the
$R_{\rm in} \propto T_{\rm in}^{-1}$ property, and that this partially
explains the too-high values of $T_{\rm in}$.
However, we at the same time presume that the issue cannot be fully
solved even assuming the slim disk hypothesis.

\subsection{Spinning BH scenario}
\label{sec:spinning BH scenario}

Before the present paper, some authors
%(e.g., Mizuno et al.\ 1999; Makishima et al.\ 2000)
\citep{Mizuno1999, Makishima2000}
tried to explain the ``too-high $T_{\rm in}$'' of ULXs by
assuming that the central BH is rapidly rotating;
this ``Kerr-BHB hypothesis'' has first been proposed by
%Zhang et al.\ (1997)
\citet{Zhang1997}
to explain the observed high temperatures
of Galactic jet sources GRS~1915+105 and GRO~J1655-40.
In this subsection, we examine whether the remaining inconsistency
seen for our ULXs can be solved by considering Kerr BHs or not,
taking into account the relativistic effects that have not
been considered by
%Makishima et al.\ (2000).
\citet{Makishima2000}.
Hereafter, we express the BH angular momentum $J$ in a dimensionless
manner,
i.e., by a spin parameter $a^{*} \equiv \frac{c}{G M^{2}}J$.
This parameter takes values between $-1$ to $1$;
$a_{*}=1$ means the extreme Kerr hole for a prograde disk
(i.e., rotating in the same direction as the BH),
$a_{*}=-1$ also represents the extreme Kerr hole but for a retrograde
disk, and $a_{*}=0$ corresponds, of course,
to the Schwarzschild BH.

The most immediate effect of the BH spin is
that it affects the radius of the last stable orbit,
$R_{\rm last}$.
While $R_{\rm last}$ is $3 R_{\rm S}$ for a
Schwartzshild BH, it reduces down to $\frac{1}{2} R_{\rm S}$
for a prograde disk around an extreme Kerr hole
of $a^{*}=1$
%(e.g. Berdeen et al.\ 1972).
\citep{Berdeen1972}.
A smaller $R_{\rm in}$ leads to a higher $T_{\rm in}$,
suggesting that the BH spin can explain the problem with ULXs.

Neglecting relativistic effects for the moment,
we can perform simple quantitative estimates.
For an extremely Kerr hole, $R_{\rm in}$ decreases by a factor of 6,
or $T_{\rm in}$ at the luminosity maximum increases
by a factor of $\sqrt{6}$.
Even taking a somewhat less extreme case of $a_{*}=0.95$,
we expect $R_{\rm in}$ to decrease
by a factor of 3 
(or, $T_{\rm in}$ increases by a factor of $\sqrt{3}$)
compared to the case of a Schwarzschild BH.
This is apparently sufficient to explain the observed high temperatures
of ULXs.

\begin{figure}
%\plotone{../summary/zhang2.ps}
%\plotone{zhang2.ps}
\plotone{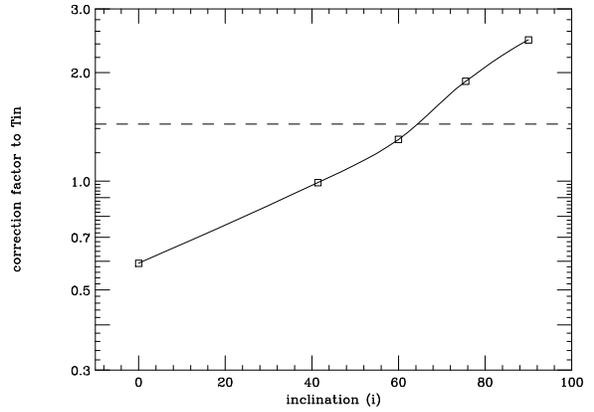}
\caption{Expected increment factor of the disk temperature
due to a BH rotation, relative to the Schwartzschild BH of the same 
mass and the same mass accretion rate.
It was calculated 
based on the correction factor of
Zhang et al (1997),
for the accretion disk around a Kerr hole with $a^{*}=0.9981$.
This value corresponds to the case where a BH equilibrates with the
disk accretion (Throne\ 1974).
The disk is assumed to be in the regime of standard accretion disk,
with no advection effects.
The dashed horizontal line indicates the goal necessary to 
explain the high $T_{\rm in}$ of ULXs.
\label{fig:zhang2.ps}
}
\end{figure}

Of course, the
X-ray spectra emergent from an accretion disk
is subject to several relativistic corrections,
due to gravitational redshift,
transverse and longitudinal Doppler shifts,
and gravitational focusing.
To a distant observer,
both the observed color temperature and flux
will deviate from the local values,
depending on the inclination angle $i$.
These effects have been examined through numerical calculations
by several authors,
%including Cunningham (1975), Asaoka (1989),
%and Zhang et al. (1997).
including \citet{Cunningham1975}, \citet{Asaoka1989},
and \citet{Zhang1997}.
Employing the numerical calculation by
%Cunningham (1975), Zhang et al. (1997)
\citet{Cunningham1975}, \citet{Zhang1997}
represent the relativistic effects by two correction
factors, which are to be applied on the MCD modeling of 
the accretion disk around a Kerr BH.
One correction factor is the change of the color
temperature denoted $\kappa_{\rm GR}$,
due to the gravitational redshift and Doppler red/blue-shifts.
The other is the change of the flux denoted $g$,
due to the viewing geometry and the gravitational focusing.
The observed color temperature scales as
$\propto \kappa_{\rm GR}$, whereas the observed flux scales as
$f_{\rm X} \propto \frac{1}{g}$ instead of
$f_{\rm X} \propto \frac{1}{\cos i}$ in the Newtonian case,
where $i$ denotes the disk inclination angle.
The latter correction
affects the inferred upper-limit of the disk temperature
as $\propto L_{\rm bol}^{-1/4} \propto g^{1/4}$.
Further taking into account the correction
due to the decrease of $R_{\rm in}$,
which increases the maximum disk temperature
as $\propto R_{\rm in}^{-1/2}$, we calculated
the combined correction factor,
$\kappa_{\rm GR} \, g^{1/4} R_{\rm in}^{-1/2}$,
for a nearly-extreme Kerr holes
as shown in Figure~\ref{fig:zhang2.ps}.

Thus, if we assume a Kerr BH of $a_{*}$=0.998,
and view the disk from an inclination angle of $i \ge 65^{\circ}$,
the correction factor exceeds $\sqrt{2}$.
When this is combined with another factor of $\sim 1.2$
due to the slim-disk property
(\S~4.2),
we can explain the discrepancy by a factor of $\sqrt{3}$
seen for ULXs.
Since $i \sim 60^{\circ}$ is what is expected on average
when the disks are randomly oriented,
the chance probability of finding such objects is reasonably high.
We thus conclude that
the mystery of ULXs may be solved in terms of
the slim accretion disk and the BH rotation.

Of course, there remain many aspects of the
ULXs phenomena to be examined.
Observationally, an optical identification of the counter-part
is important; it can help us to understand how
ULXs are formed, and confirm whether they are really
massive, rapidly-rotating BHs.
We also need to obtain spectra with much higher quality
over a wider bandpass,
in order to compare them with the
theoretical predictions in detail.
Theoretically, numerical models of the 
X-ray spectrum emergent from an accretion disk around a
Kerr BH, taking into account both the relativistic and advection
effects, are required.
Another big theoretical issue is how to make 
massive ($\sim100~M_{\odot}$) and rapidly rotating
($a^{*} \ge 0.95$) stellar-mass BHs.
%
%Such studies are beyond the scope of this paper,
%but some of them will be performed in the near future.

\placefigure{fig:zhang2.ps}

\clearpage

\begin{deluxetable}{lcrrcc}
%\tabletypesize{\scriptsize}
\tablecaption{
$ASCA$ observational log of the host galaxies.
\label{tab:sample}
}
\tablewidth{0pt}
\tablehead{
\colhead{Galaxy} & \colhead{Date\tablenotemark{a}} &
\multicolumn{2}{c}{Exposure (ks)\tablenotemark{b}} &
\colhead{SIS Mode\tablenotemark{c}} &
\colhead{SIS Clock\tablenotemark{d}} \\
\colhead{} & \colhead{yymmdd} & \colhead{SIS} & \colhead{GIS} & 
\colhead{} & \colhead{}
}
\startdata
IC~342 & 93 09 19 & 35.8 & 38.4 & F/B  & 1111/1111 \\
NGC~3031\tablenotemark{e} & 94 04 01 & 31.6 & --  & F/F & 1000/1000 \\
(M81) & 94 10 21 & 37.4 & -- & F/F  & 0010/0010 \\
 & 95 04 01 & 17.6 & -- & F/F & 1000/1000 \\
 & 95 10 24 & 34.5 & -- & F/F & 0010/0010 \\
 & 96 04 16 & 43.2 & -- & F/F & 1000/1000 \\
 & 96 10 27 & 27.8 & -- & F/F & 0010/0010 \\
 & 97 05 08 & 41.0 & -- & F/F & 1000/1000 \\
%NGC~3031 & 94 04 01 & 31.6 & 37.6  & F/F & 1000/1000 \\
%(M81) & 94 10 21 & 37.4  & -- & F/F  & 0010/0010 \\
% & 95 04 01 & 17.6 & 19.7 & F/F & 1000/1000 \\
% & 95 10 24 & 34.5 & 37.3 & F/F & 0010/0010 \\
% & 96 04 16 & 43.2 & 47.7 & F/F & 1000/1000 \\
% & 96 10 27 & 27.8 & 31.0 & F/F & 0010/0010 \\
% & 97 05 08 & 41.0 & 48.9 & F/F & 1000/1000 \\
NGC~1313 & 93 07 12 & 21.2 & 27.7 & F/B  &
 1111/1111 \\
 & 95 11 29 & 33.2  & 32.9 & F/F & 0100/0100 \\
\enddata
\tablenotetext{a}{
Observation start date.
}
\tablenotetext{b}{
An average of the two SIS sensors or the two GIS sensors,
after the data screening is applied.
}
\tablenotetext{c}{
Data acquisition mode for Bit-High/Bit-Medium.
}
\tablenotetext{d}{
Clocking mode of S0 for Bit-High/Bit-Medium.
}
\tablenotetext{e}{
The GIS data are not utilized in the present study (see \S~3.2).
}
%\tablecomments{
%GIS bit assignment for Bit-H/Bit-M (PH-X-Y-RT-SP-Time) was
%8-8-8-0-0-7/8-8-8-0-0-7 for NGC~1313 in the 1995 observation,
%and 8-8-8-0-0-7/8-6-6-0-0-10 for M81 in the last three observations.
%All other GIS observations were conducted in the PH normal mode.
%}
\end{deluxetable}

\begin{deluxetable}{lcccc}
\tablewidth{0pt}
\tablecaption{
Time-resolved ULX spectra fitted by the MCD model.
Errors are calculated for 90\% confidence for one
interesting parameter ($\Delta \chi^2=2.7$).
\label{tab:fit_ulxs}
}
\tablehead{
\colhead{Source} & \colhead{$N_{\rm H}$} &
\colhead{$T_{\rm in}$} &
\colhead{$f_{\rm bol}^{\rm disk}$\tablenotemark{\dag}} &
\colhead{$\chi^{2}/ \nu$} \\
 & ($10^{22}~{\rm cm}^{-2}$) & (keV) & & 
}
\startdata
%NGC~1313 source~B & & & \\
% \hspace{0.5cm} 1993 Jul. & $0.08 \pm 0.04$ & 
%1.47$\pm$0.08 & 4.16 & 124.5/130 \\
% \hspace{0.5cm} 1995 Nov. & $0.06 \pm 0.05$ &
%1.07$\pm$0.07 & 1.66 & 89.8/74 \\ \hline
\sidehead{IC~342 source~1}
% \hspace{0.5cm} time-average & $0.47 \pm 0.03$ &
%1.77$\pm$0.05 & 13.7 & 137.4/135 \\
high-flux & $0.47 \pm 0.04$ &
1.87$\pm$0.07 & 15.5 & 129.0/106 \\
low-flux & $0.47 \pm 0.07$ &
1.42$\pm$0.09 & 8.95 & 115.6/107 \\
% \hspace{0.5cm} phase 1 & 0.47(fix) &
%1.96$\pm$0.10 & 16.2 & 59.2/81 \\
% \hspace{0.5cm} phase 2 & 0.47(fix) &
%1.50$\pm$0.10 & 10.3 & 60.4/63 \\
% \hspace{0.5cm} phase 3 & 0.47(fix) &
%1.70$\pm$0.15 & 13.0 & 52.2/48 \\
% \hspace{0.5cm} phase 4 & 0.47(fix) &
%1.29$\pm$0.08 & 7.95 & 54.3/60 \\
% \hspace{0.5cm} phase 5 & 0.47(fix) &
%1.81$\pm$0.07 & 15.1 & 137.4/135 \\
\sidehead{M81~X-6}
1994 April & $0.19 \pm 0.06$ &
1.63$\pm$0.19 & 3.61 & 43.8/42 \\
1994 October & $0.23 \pm 0.06$ &
1.51$\pm$0.14 & 4.05 & 34.9/47 \\
1995 April & $0.14^{+0.15}_{-0.11}$ &
1.73$\pm$0.27 & 3.71 & 29.6/23 \\
1995 October & $0.08 \pm 0.08$ &
1.34$^{+0.23}_{-0.18}$ & 2.48 & 17.7/21 \\
1996 April & $0.33^{+0.18}_{-0.14}$ &
1.23$^{+0.18}_{-0.15}$ & 2.78 & 61.6/54 \\
1996 October & $0.08 \pm 0.07$ &
1.63$^{+0.23}_{-0.19}$ & 4.14 & 30.2/31 \\
1997 May. & $0.10 \pm 0.07$ & 
1.61$^{+0.19}_{-0.16}$ & 3.80 & 44.5/40 \\
high-temp phase & $0.16 \pm 0.03$ &
1.59$\pm$0.09 & 3.92 & 68.6/65 \\
low-temp phase & $0.20 \pm 0.08$ &
1.29$\pm$0.13 & 2.63 & 89.8/74 \\
\enddata
\tablenotetext{\dag}{
Bolometric flux of the accretion disk in units of
10$^{-12}~{\rm erg}~{\rm s^{-1}}~{\rm cm^{-2}}$,
assuming a face-on geometry.
}
\end{deluxetable}

\end{document}